\documentclass[reprint,english,superscriptaddress,preprintnumbers,amsmath,amssymb,aps,prx]{revtex4-1}

\usepackage{graphicx}
\usepackage{subfigure}
\usepackage{dcolumn}
\usepackage{bm}
\usepackage[mathlines]{lineno}

\usepackage{xspace}
\usepackage{siunitx}
\usepackage{hyperref}
\usepackage[nameinlink]{cleveref}
\usepackage{appendix}

\usepackage{changes}

\def\eq#1{eq.~\eqref{#1}}

\def\fig#1{Fig.~\ref{#1}}

\def\tab#1{Table~\ref{#1}}

\graphicspath{
	{./figures/}
	{../figures/}
}

\newcommand{\imag}{\textrm{i}}
\newcommand{\spatial}[1]{\vec{#1}}

\def\intsum{\int_n\llap{$\sum\,$}}

\newcommand{\re}{{\textrm{Re}}}
\newcommand{\im}{{\textrm{Im}}}

\newcommand{\gev}{\textrm{GeV}}

\newcommand{\lye}{\text{LYE}}

\usepackage{xcolor}

\begin{document}

\title{Lee Yang edge singularities of QCD in association with Roberge-Weiss phase transition and chiral phase transition}

\author{Zi-yan Wan}
\email[]{wanziyan@stu.pku.edu.cn}
\affiliation{Department of Physics and State Key Laboratory of Nuclear Physics and Technology, Peking University, Beijing 100871, China}

\author{Yi Lu}
\email[]{qwertylou@pku.edu.cn}
\affiliation{Department of Physics and State Key Laboratory of Nuclear Physics and Technology, Peking University, Beijing 100871, China}

\author{Fei Gao }
\email{fei.gao@bit.edu.cn}
\affiliation{School of Physics, Beijing Institute of Technology, 100081 Beijing, China}

\author{Yu-xin Liu }
\email{yxliu@pku.edu.cn}
 \affiliation{Center for High Energy Physics, Peking
University, Beijing 100871, China}
\affiliation{Department of Physics and State Key Laboratory of
Nuclear Physics and Technology, Peking University, Beijing 100871,
China}
\date{\today}

\begin{abstract}
We study the Quantum Chromodynamics (QCD) phase transitions in the complex chemical potential plane in the framework of Dyson-Schwinger equation approach, in the presence of a constant gluonic background field that represents confining dynamics. We solve the quark gap equation and the background field equation self consistently, which  allows us to directly explore the confinement phase transition and furthermore, evaluate the impact of the back-coupling of confinement on chiral symmetry breaking. Moreover, within such a coupled framework towards the complex chemical potential region, we demonstrate the emergence of Roberge-Weiss (RW) symmetry and investigate the trajectory of Lee-Yang edge singularities (LYES). Our analysis reveals that the LYES scaling behavior is  similar to our previous findings without the background field condensate~\cite{Wan:2024xeu}. However, a significant difference from our earlier work is that the trajectory of LYES terminates when the imaginary part of the singularity becomes $1/3 \, \pi T$. We elaborate that this cut-off behavior is caused by the RW symmetry that is symmetric to the imaginary chemical potential $\mu_i=1/3 \, \pi T$.

\end{abstract}

\maketitle

\section{Introduction}
The study of QCD (Quantum Chromodynamics) phase transitions encompasses two important aspects: the chiral phase transition and the deconfinement phase transition. Considerable efforts have been devoted to elucidating the properties of these transitions and their interrelations~\cite{Aarts:2023vsf,Fu:2022gou,Fischer:2018sdj,Schaefer:2008ax}. 
For the chiral phase transition, there is clear definition for its order parameter, i.e., the chiral condensate~\cite{Borsanyi:2025lim,Philipsen:2021qji,HotQCD:2019xnw,Gao:2020qsj,Gao:2020fbl,Fu:2019hdw,Gunkel:2021oya,Braun:2023qak}, while the confinement remains less definitive. 
Various criteria for confinement have been proposed, see reviews in e.g.
\cite{Fischer:2006ub,Greensite:2011zz,Reinosa:2024njc,Dupuis:2020fhh}. 
%
Here, we focus on the Polyakov loop, 
more precisely the gluonic background field~\cite{Braun:2007bx,Fister:2013bh,Reinosa:2014ooa}, which stems from the thermal analog of the Wilson loop and serves as a crucial candidate of the order parameter on quark confinement. 
%

The interplay between these two transitions provides valuable insights into the complex dynamics of strong interactions within the context of quantum chromodynamics (QCD). Moreover, a new approach for studying QCD phase transitions has been proposed based on the Lee-Yang edge singulairties (LYES)~\cite{Connelly:2020gwa,Rennecke:2022ohx,Johnson:2022cqv,Ihssen:2022xjv,Zambello:2023ptp,Clarke:2024seq,Adam:2025pii}, which requires one to extend the calculation into the imaginary chemical potential region.  The imaginary chemical potential itself induces a more complex relation between the chiral and deconfinement phase transition, because a new symmetry, the  Roberge-Weiss (RW) symmetry becomes evident at non-zero imaginary chemical potential. There further exists  the Roberge-Weiss transition~\cite{Roberge:1986mm}, which directly has impact on the LYES.  Numerous studies on this topic have been conducted by using various methods \cite{Sakai:2008py, Kashiwa:2011td, Fischer:2014vxa, Cuteri:2022vwk}, and including investigations into its LYES \cite{Dimopoulos:2021vrk}. Additionally, the results presented later reveal that RW symmetry can partially influence the behavior of the chiral LYES trajectory discussed in our previous work \cite{Wan:2024xeu}.

To incorporate the confinement aspect into the current study, we utilize the gluonic background field ($\bar{A}_4$) approach in functional QCD. Such an approach has been developed both in the Dyson-Schwinger equations (DSE)~\cite{Fister:2013bh, Fischer:2013eca, Fischer:2014ata, Fischer:2014vxa, Lu:2025cls} and in the functional renormalisation group (fRG)~\cite{Braun:2007bx, Braun:2009gm, Braun:2010cy, Marhauser:2008fz, Fister:2013bh, Herbst:2015ona}, which has also been further applied in e.g. the Curci-Ferrari model~\cite{Reinosa:2014ooa, Reinosa:2015oua, Maelger:2017amh, 10.21468/SciPostPhys.12.3.087, vanEgmond:2024ljf, MariSurkau:2025pfl}. 
Specifically, we shall focus on the DSE approach in the present study, see~\cite{Roberts:1994dr,Alkofer:2000wg, Roberts:2000aa,Fischer:2006ub,Fischer:2018sdj} for its applications on the chiral symmetry aspect. 
As for the confinement aspect, the key is to evaluate the quantum equation of motion of $A_4$ from the effective potential $\Gamma$, namely
%
%
%
$\frac{\delta \Gamma}{\delta A_4}  = 0$, which would naturally yield a nonzero $\bar{A}_4$ in the confining regime~\cite{Fister:2013bh,Fischer:2013eca}. 
Further applications on RW symmetry can also be found in~\cite{Fischer:2014vxa}. 
%
In fact, recent studies have also revealed the importance of the $A_4$ condensate on thermodynamic quantities, especially on the conserved charge fluctuations~\cite{Lu:2025cls, Rennecke:2019dxt}. 
%

Here in the present work, we further determine $\bar{A}_4$ self-consistently via the coupled solution between the chiral and confinement dynamics
in \Cref{model}. 
Then in \Cref{sec:sym}, we illustrate the symmetries observed in the background field and the Polyakov loop potential, including the center symmetry and the RW symmetry. In \Cref{results}, we show results on QCD phase transitions and the RW transition in the complex chemical potential plane, which also allows us to extract the respective Lee-Yang edge singularities. 
Finally, in \Cref{sum} we summarise and make conclusions.
%
%


\section{DSEs with background field and Polyakov loop potential}\label{model}
The Polyakov loop, first mentioned by \cite{Polyakov:1978vu}, represents the free energy of a single static quark, given by the expression $L = e^{-F_q/T}$. When $L = 0$, it implies an infinite free energy $F_q$, suggesting the impossibility of a single quark's existence—a manifestation of confinement. Additionally, $L = 0$ signifies the preservation of center symmetry\cite{Svetitsky:1982gs}, which can be mathematically expressed through the invariance of $L$ under center transformations: $L = zL$, where $z$ is an element of the center of $\mathrm{SU}(N_c)$ gauge group: $\mathrm{Z}(N_c)$. Considerable efforts have been directed towards exploring confinement by $L$ using lattice computations\cite{Bazavov:2011nk} and investigating the effects of the chiral order parameter's coupling to the Polyakov loop through models\cite{Meisinger:1995ih,Schaefer:2007pw}. 
 we utilize a derived order parameter from Polyakov loop, namely the Polyakov loop potential. The potential involves computing Polyakov loop $L(\langle A_4\rangle)$ by incorporating a background gluon field  condensate as $A_4=\bar{A}_4$~\cite{Marhauser:2008fz,Braun:2007bx}.
 Built on these methods, here we would like to incorporate the gluonic background field in the DSEs. 
\subsection{Gluonic background field and DSEs}
The DSEs involving a constant background field $\bar{A}_4$ have been discussed extensively in \cite{Fister:2013bh, Fischer:2014vxa, Fischer:2013eca,Lu:2025cls}. Based on these studies, we demonstrate for the first time a fully coupled, self-consistent resolution of $A_4$ within the DSEs truncated up to 2-point functions. Such a framework is illustrated below in detail.

With $\bar{A}_\mu$, the total gluon field is given by $A_\mu = a_\mu + \bar{A}_\mu$, where $a_\mu$ represents the fluctuating quantum field, serving as the integration variable in the functional integral.  To simplify the Lagrangian's formulation, the original derivative $\partial_\mu$ can be replaced by the background-field covariant derivative: $\bar{D}_\mu =\partial_\mu-ig\bar{A}_\mu$, and the corresponding adjoint form $\bar{D}_\mu^{ab} = \partial_\mu \delta^{ab} + gf^{acb}\bar{A}_\mu^c$. Further details can be found in \cite{Abbott:1981ke, Peskin:1995ev}. The final form of the Lagrangian is: 
\begin{equation}\label{eq:Lagrangian}
	\begin{aligned}
		\bar{\mathcal{L}}= & \bar{\psi}(i\bar{D}_\mu \gamma^\mu+g a_\mu\gamma^\mu-m)\psi-\frac{1}{4}\bar{F}^a_{\mu\nu}\bar{F}_a^{\mu\nu}\\
		& -\frac{1}{2 \xi}(\bar{D}^\mu a^a_\mu)^2-\bar{c}^a\bar{D}^{ac}_\mu[ \bar{D}^{cd}_\mu -g f^{cdf}a_\mu^f] c^d
	\end{aligned}
\end{equation}
We will take $\xi=0$, the Landau-deWitt gauge, which requests $\bar{D}^\mu a^a_\mu=0$.

After following the derivation process outlined in \cite{Roberts:2000aa}, the DSE for the quark propagator in the presence of a background field can be derived. Here, we present the  {DSE} for finite temperature $T$ and chemical potential $\mu$, where the $O(4)$ symmetry of the momentum $p$ is broken down to $O(3)$ (with $\spatial{p}$ and $\omega_{n}$). Specifically, $\omega_{n} = \pi T(2n+1)$ represents the Matsubara frequency for fermions:
\begin{equation}\label{DSE}
	\begin{split}
		& \left[\bar{S}(p;\bar{A}_4)\right]^{-1} = \left[\bar{S}_0(p;\bar{A}_4)\right]^{-1} + \bar{\Sigma}(p), \\
		& \bar{\Sigma}(p) = Z_{1 } g^{2} T \intsum \frac{d^{3} \spatial{\,q}}{(2 \pi)^{3}} \gamma_{\mu}t^a \bar{S}(q) \bar{D}^{ab}_{\mu \nu}(k;\bar{A}_4) \bar{\Gamma}^b_{\nu}(p,q;k) \, .
	\end{split}
\end{equation}
The dressed gluon propagator $\bar{D}_{\mu\nu}^{ab}(k;\bar{A}_4)$ and the gluon-quark vertex $\bar{\Gamma}_{\nu}^{b}(p, q;k)$ are both defined within the context of a background gluon field. Additionally, the expression for the renormalized bare quark propagator  $\bar{S}_0(p;\bar{A}_4)$ is given by:
\begin{equation}\label{eq:Sq0}
\bar{S}_0(p;\bar{A}_0)^{-1}  =  \imag Z_{2} \left(\gamma_{4}\tilde{\omega}_{n}+\spatial{p}\spatial{\gamma}\right) + Z_{2}Z_{m} m_{f} \, ,
\end{equation}
with abbreviation $\tilde{\omega}_{n} = \omega_{n} + i\mu + g\bar{A}_{4}$. $Z_{2}$, $Z_{1}$, and $Z_{m}$ represent the renormalization constants for the quark wavefunction, vertex, and quark mass, respectively. And $g^{2} = 4\pi \alpha(\zeta)$ denotes the coupling constant, which depends on the renormalization scale $\zeta$. From Equation \ref{eq:Sq0}, the influence of the background field $\bar{A}_4$ can be regarded as a momentum shift in the Matsubara frequency for each color component $i, j$:
\begin{equation}\label{eq:Sq0A}
	[\bar{S}_{0}(\omega_{n},\spatial{p};\bar{A}_4)^{-1}]_{ij} = [\bar{S}_{0}(\omega_{n}+g \bar{A}_4^a t^a_{ij},\spatial{p})^{-1}]_{ij},
\end{equation}
%
The background field $\bar{A}_4$ can be diagonalized into the Cartan sub-algebra:
\begin{equation}\label{eq:A4}
   \bar{A}_4=\frac{1}{g} 2 \pi T \, (\phi_3 t^3+\phi_8 t^8),
\end{equation}  
with $t^a=1/2 \lambda^a$, $\lambda^a$ is the Gell-Mann Matrix, and the Polyakov loop $L(\bar{A}_4)$ can be derived as:
\begin{equation}\label{eq:L}
	L\left(\bar{A}_4\right)=\frac{1}{3}\left[e^{-i \frac{2 \pi \phi_8}{\sqrt{3}}}+2 e^{i \frac{\pi \phi_8}{\sqrt{3}}} \cos \left(\pi \phi_3\right)\right].
\end{equation}
Therefore, the bare quark propagator possesses only a diagonal element, given by $\bar{S}_{0,ij} = \bar{S}_{0,ii} \delta_{ij}$. For $i \neq j$, the quark propagator $S^{-1}(p)$ diminishes towards zero as $p \rightarrow \infty$, a consequence of \textit{Asymptotic Freedom}. For simplicity, we neglect the non-diagonal term across all momentum ranges, 
thereby expressing the dressed quark propagators as:
\begin{equation}\label{eq:dressed_quark_propagator}
	\left[\bar{S}(p)\right]^{-1}_{ij}\!\! \!= \delta_{ij}[ \imag \tilde{\omega}_{n} \gamma_{4} C_i(\vec{p},\tilde{\omega}_{n})+ \imag \vec{p} \cdot \vec{\gamma} A_i(\vec{p},\tilde{\omega}_{n}) + B_i(\vec{p},\tilde{\omega}_{n})].
\end{equation}
 Furthermore, the quark Dyson-Schwinger equations (DSEs) can be decomposed into coupled equations pertaining to the three color components of the quark. 

\begin{figure}[htbp]
	\centering
	\includegraphics[width=0.75 \linewidth]{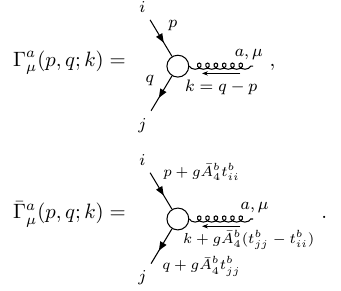}
	\caption{Upper panel: Feynman diagram depicts the momentum flow for each leg of quark-gluon vertex without the influence of a background field. Lower panel: the diagram showcases the effects of $\bar{A}_4$ on the quark leg, carrying color indices $i$ and $j$, and the gluon leg, indexed by $a$, where these effects are treated as momentum translations. This figure and the subsequent one \fig{fig:polya} are generated by Refs.~\cite{Ellis:2016jkw,Dohse:2018vqo}.
    }  \label{fig:vtx}
\end{figure}

Analogously, considering the background field, the bare gluon propagator can be derived as:
\begin{equation}\label{eq:bare_gluon_propagator}
	[D_0^{-1}]_{ab}^{\mu\nu} = z_{a} \left[\left(\frac{1}{\zeta} - 1\right)\bar{D}_{\mu}^{ca}\bar{D}_{\nu}^{cb} + \delta_{\mu\nu}\bar{D}_{\lambda}^{ca}\bar{D}_{\lambda}^{cb}\right],
\end{equation}
where $z_{a}$ represents the gluon wave function renormalization, and $\bar{D}_{\mu}^{ab}$ is the adjoint covariant derivative: $\partial_{\mu}\delta_{ab} + gf_{acb}\bar{A}_{\mu}^{c}$, as discussed previously. The second term contributes to the non-diagonal components. However, when acting on $t^{b}$, this term transforms into $k^{\mu}t_{a} + g[t_{a}, t_{c}]\bar{A}_{4}^{c}$, modifying the original $k^{\mu}$, the coefficient of non-zero $(t_{a})_{ij}$, to $k^{\mu} + g\bar{A}_{4}^{c}(t_{c}^{jj} - t_{c}^{ii})$ for $c = 3, 8$. The introduction of the background field does not convert the original zero color component into a non-zero one. Consequently, we adopt a scheme that considers the impact of the recent background field $\bar{A}_{4}$ on propagators and the gluon-quark vertex as a momentum translation, illustrated in \fig{fig:vtx}.
Now, the quark DSE can be simplified to following coupled equations, for $i=\{1,2,3\}$:
\begin{equation}
	\begin{split}
		\bar{S}_{ii}^{-1}(p\!+\!g\bar{A}^4_{ii})  =  \bar{S}^{-1}_{0,ii}(p\!+\!g\bar{A}^4_{ii})+Z_{1 } g^{2} T \intsum \frac{d^{3} \spatial{\,q}}{(2 \pi)^{3}}\cdot&\\
		  \sum_j\left[\gamma_{\mu}t_{ij}^at_{ji}^a \bar{S}(q\!+\!g\bar{A}^4_{jj}) D_{\mu \nu}(k\!+\!g\bar{A}^4_{jj}-g\bar{A}^4_{ii})\Gamma_{\nu}\right].&
	\end{split}
\end{equation}
For the dressed gluon propagator $D^{\mu\nu}$ and vertex $\Gamma_\nu$, in order to compare the influence of the background field on both the chiral phase transition and Lee-Yang edge singularities, we adopt the truncation scheme as in our previous work, Ref.~\cite{Wan:2024xeu}. Specifically, we employ the "functional-lattice" gluon propagator given by $D_{\mu \nu}(p) = D(p) P_{\mu \nu}$, where $P_{\mu \nu}$ denotes the corresponding Lorentz structure, as in Ref.~\cite{Gao:2021wun}: 
\begin{equation}\label{eq:gln}
	\begin{split}
		&D(p)=\frac{\left(a^{2}+p^{2}\right)/\left(b^{2}+p^{2}\right)}{M^{2}\left(p^{2}\right)+p^{2}\left[1+c\ln\left(d^{2}p^{2}+e^{2}M_G^{2}\left(p^{2}\right)\right)\right]^{\gamma}},\\
		&\text{ with }	M_G^{2}(p)=\frac{f^{4}}{g^{2}+p^{2}},
	\end{split}
\end{equation}
and the gluon-quark vertex in Refs.~\cite{Fischer:2014ata,Gunkel:2021oya}:
\begin{equation}\label{vertex}
	\begin{split}
		& \Gamma_{\mu}(p,q;k) =  \gamma_{\mu}\Gamma(k^{2})\! \left[\delta_{\mu4}\frac{C(p)\!+\!C(q)}{2}\!+\!\delta_{\mu i}\frac{A(p)\!+\!A(q)}{2}\right]\\
		& \Gamma\left(k^{2}\right) =  \frac{d_{1}}{d_{2}\!+\! k^{2}} \! + \! \frac{k^{2}}{\Lambda^{2}\! + \! k^{2}}\! \Bigg{(}\! \frac{\beta_{0}\alpha(\zeta)\ln\left[k^{2}/\Lambda^{2}+1\right]}{4\pi}\! \Bigg{)}^{-2\delta}.
	\end{split}
\end{equation}
The anomalous dimensions are $\gamma=(13 - 4/3N_{f})/(22 - 4/3N_{f})$, $\beta_{0}=(11N_{c} - 2N_{f})/3$, $\delta=9N_{c}/(44N_{c} - 8N_{f})$. Here, we take $N_f=2+1,N_c=3$. Because a constant background field does not affect the vacuum solution, the parameters in the truncation scheme determined by vacuum physical values remain the same as listed in Ref.~\cite{Wan:2024xeu}. 
Therefore, the other parameters are $\{a,b,c,d,e\}=\left\{ 1\,\mathrm{GeV},\, 0.735\,\mathrm{GeV},\, 0.12,\, 0.0257\,\mathrm{GeV}^{-1},\, 0.081\,\mathrm{GeV}^{-1}\right\}$, together with $\{f,g\}=\{0.65\,\mathrm{GeV},\, 0.87\,\mathrm{GeV}\}$, $\alpha(\zeta)=0.3$, $\Lambda = 1.4\,\textrm{GeV}$, $d_{1}=6.8\,\textrm{GeV}^{2}$ and $d_{2} = 0.5\,\textrm{GeV}^{2}$.

\begin{figure}[htbp]
	\includegraphics[width=0.9 \linewidth]{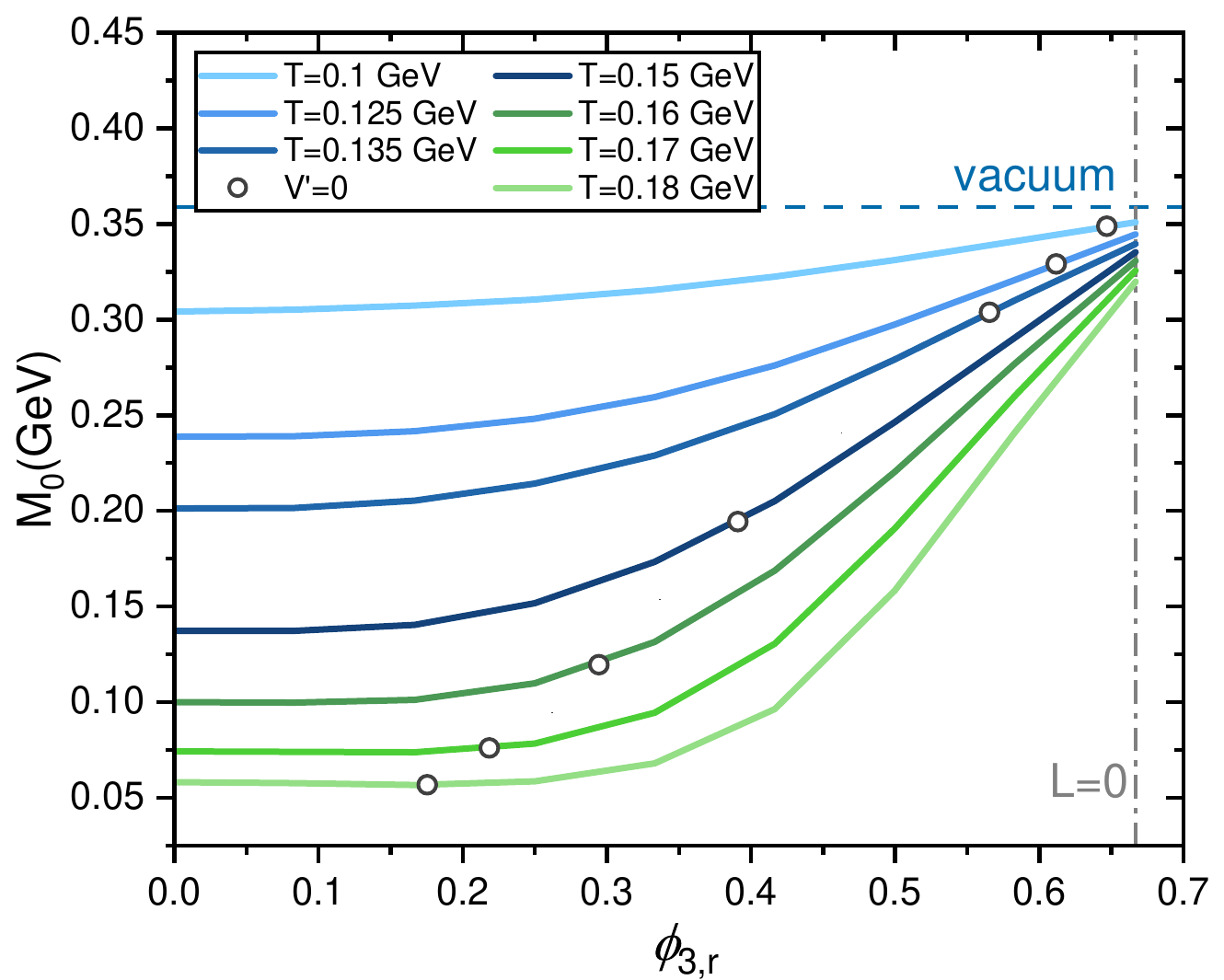}
	\caption{The third color component of the dressed quark mass $M_0 = M(\mathbf{0}, \omega_1)$  with respect to $\phi_{3,r}$ at various temperatures $T$. The vacuum value is denoted by a dashed blue line. $L=0$, specifically when $\phi_{3,r} = 2/3$, is indicated by a dash-dot gray line. And the black dots signify the points at which the Polyakov loop potential attains its minimum value.}
	\label{fig:Mlist}
\end{figure}

Then, we can solve the DSEs with an arbitrary constant background field $\bar{A}_4$. \fig{fig:Mlist} shows the variation of the third color component of the dressed quark mass $M(\spatial{0},\omega_1)$ with respect to a real $\phi_{3,r}$ at different temperatures $T$. Obviously, a non-zero $\phi_{3,r}$ enhances the strength of the dressed quark mass. When $\phi_{3,r}$ reaches $2/3$, where the Polyakov loop $L = 0$, the values of the dressed mass at various temperatures approach close to the vacuum value. The diagram illustrates how confinement effects influence chiral symmetry.
\subsection{Polyakov loop potential}
%
The physical value of the gluonic background field
corresponds to the minimum point of the Polyakov loop potential $V(\bar{A}_4)$. The DSE for the background gluon one-point function, i.e. the derivative $\text{d}V/\text{d}\bar{A}_4$,  has been derived in \cite{Fister:2013bh, Fischer:2013eca}. 
{Here, we analyze exclusively the one-loop term in \fig{fig:polya}, whose physical viability is demonstrated in \cite{Fister:2013bh}}. The specific formula for this one-loop term of the derivative of the potential with respect to $\phi$: $V'=\text{d} V/\text{d}\phi$, given $\bar{A}_4 = \frac{2\pi T}{g}\phi \, \text{diag}(\nu_l, \{l\})$, where $\nu_l$ are the eigenvalues of the generating element of $\bar{A}_4$ in the fundamental ($\nu_{ql}$) or adjoint ($\nu_{gl}$) representation in QCD, is:
\begin{figure}[htbp]
	\includegraphics[width=0.7 \linewidth]{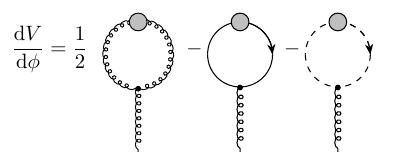}
	\caption{One loop truncated DSE for the background gluon one-point function.}
	\label{fig:polya}
\end{figure}
\begin{equation}\label{eq:Polya}
	\begin{split}
		&V'\!=\!\intsum\!\left[\frac{1}{2}\sum_{gl}\nu_{gl}\tilde{\omega}_{n,gl}\!\cdot\!\left(z_{a}G_{L}\!(\spatial{q},\tilde{\omega}_{n,gl})\!+\!2z_{a}G_{T}(\spatial{q},\tilde{\omega}_{n,gl})\right.\right.\\
		&\left.\left.-2z_{c}G_{c}(\spatial{q},\tilde{\omega}_{n,gl})\right)\!-\!\!\!\sum_{ql,f}\nu_{ql}\tilde{\omega}_{n,ql}Z_{2f}S_{f}(\spatial{q},\tilde{\omega}_{n,ql})\right]\!\!+\!\frac{1}{2}V'_{\mathrm{Weiss}}
	\end{split}
\end{equation}
with the abbreviated boson and fermion Mastubara frequency given by:
\begin{equation}
\begin{split}
&\tilde{\omega}_{n,gl}=2\pi T(n+\phi\nu_{gl});\\ &\tilde{\omega}_{n,ql}=\pi T(2n+1+2\phi\nu_{ql})+i\mu.
\end{split}
\end{equation}
And $\frac{1}{2}V_{\mathrm{Weiss}}$ is the perturbative one-loop potential as \cite{Weiss:1980rj} in 4 dimensions, 
\begin{equation}
	V_{\mathrm{Weiss}}=-\frac{2}{\pi^{2}} T^4 \sum_l^{N_c^2-1} \sum_{n=1}^{\infty} \frac{\cos \left\{2 \pi n \nu_l|\varphi|\right\}}{n^4},
\end{equation} 
The product of the renormalization constant with the propagator, $z_x G_x = z_x \cdot \frac{1}{Z_x(p^2)} p^2$, is renormalization invariant. As the renormalization point $\zeta$ changes to $\zeta'$, the transition of the renormalized propagator and renormalization constant is given by:
\begin{equation}
	\begin{split}
		G_x(p;\zeta) = \frac{Z_x(\zeta^2)}{Z_x(p^2) p^2} \; &\text{to} \; G_x(p;\zeta') = G_x(p;\zeta) \frac{Z_x(\zeta'^2)}{Z_x(\zeta^2)}, \\
		z_x(\zeta) \qquad &\text{to} \quad z_x(\zeta') = z_x(\zeta) \frac{Z_x(\zeta^2)}{Z_x(\zeta'^2)}.
	\end{split}
\end{equation}
This guarantees the renormalization invariance of the DSE: $G^{-1}_x(\zeta') = z_x(\zeta')G^{-1}_{0,x} + \Sigma(\zeta')$. Consequently, one can compute \eq{eq:Polya} at the same point $\zeta = 40\,\gev$ as we solve the DSEs. For better convergence of the sum of Matsubara frequencies and to ensure renormalization invariance, choose $Z_{2f} = C(p)$ in \eq{eq:Polya}. Additionally, 
It is found in Ref.~\cite{Gao:2020fbl} that the Euclidean $O(4)$-symmetry is effectively restored for high Matsubara frequencies, where the dressing functions converge rapidly towards their vacuum counterparts. To that end, 
when $\sqrt{\mathbf{p}^2+\text{Re}(\tilde{\omega}_n)^2} > 3\pi T$, we take $O(4)$ approximation for $A(\omega_n, \mathbf{p})$ and $C(\omega_n, \mathbf{p})$, which are replaced by a (3,3) Padé fit function of the vacuum $A(p)$. We take the same form gluon propagator as \eq{eq:gln} in Polyako-loop computation. As for the ghost propagator $G_c(k^2) = F(k^2)/k^2$, where $F(k^2)$ is the ghost dressing function, it is taken as described in Refs.~\cite{Cyrol:2017ewj, Zheng:2023tbv}.
\begin{equation}
	F\left(k^2\right)=\frac{\left(a_1+b_1 \sqrt{k^2}+k^2\right) /\left(c_1+d_1 \sqrt{k^2}+k^2\right)}{\left[1+e_1 \ln \left(f_1^2 k^2+g_1^2 M_G^2\left(k^2\right)\right)\right]^\delta}
\end{equation}
where $M_G^2\left(p^2\right)$ has been defined in \eq{eq:gln}. And the other fit parameters are $\left\{a_1, b_1, c_1, d_1, e_1\right\}= \left\{0.152 \gev^2, 0.697 \gev, 0.0055 \gev^2, 0.016 \gev, 0.045\right\}$, and $ \left\{f_1, g_1\right\}=\left\{0.025 \gev^{-2},0.0237\gev^{-2}\right\}$.

 Next step is using \eq{eq:Polya} to find the minimum points $(\bar{A}_4,L(\bar{A}_4))$ and substitute the values back into DSEs.
\section{A general symmetry analysis  of background field and Polyakov loop potential}\label{sec:sym}
Before presenting further Dyson-Schwinger equation (DSE) results, we first discuss the symmetry of the background condensate in $(\phi_3,\phi_8)$ plane for a non-zero chemical potential $\mu = \mu_r - i\pi T\theta$. The symmetry analysis helps to simplify the process of finding the minimum points of the Polyakov loop potential  as discussed in \cite{Reinosa:2015oua}, besides,  it  also provides insights into center symmetry and Roberge-Weiss symmetry.

\subsection{Center Symmetry}\label{sec:center_sym}

As mentioned in the previous section \ref{model}, the effect of $\bar{A}_4$ on the quark propagator can be interpreted as a shift of the Matsubara frequencies $\omega_n$:
\begin{equation}\label{eq:wn}
	\begin{aligned}
		&\omega_n\rightarrow \omega_n+i \mu_r+\theta \pi T+\phi_3 \pi T-\phi_8 \frac{1}{\sqrt{3}} \pi T \\
		& \omega_n\rightarrow \omega_n+i \mu_r+\theta\pi T-\phi_3 \pi T-\phi_8 \frac{1}{\sqrt{3}} \pi T \\
		& \omega_n\rightarrow \omega_n+i \mu_r+\theta \pi T+\phi_8 \frac{2}{\sqrt{3}} \pi T
	\end{aligned}
\end{equation}
These equations illustrate the symmetries within the $(\phi_3,\phi_8)$ plane,  indicating that the solution to the DSEs remains invariant under certain transformations. First of all, it is straightforward to conclude that the real chemical potential does not change the symmetry. The symmetries are summarized as follows: 

\indent (a) the real parts of $\phi_3$ and $\phi_8$ undergo translation periodic symmetry with periods of $2\pi$ and $2\sqrt{3}\pi$, respectively:
\begin{equation}\label{eq:symt}
	\begin{split}
		&\phi_{3,r}\rightarrow\phi_{3,r}+2\pi n,\\
		&\phi_{8,r}\rightarrow\phi_{8,r}+2\sqrt{3}\pi m, \ (n,m\in \mathbb{Z}).
	\end{split}
\end{equation}
\indent (b) The real parts of $\phi_3$ and $\phi_8$ take half-period translation symmetry when shifted simultaneously:

\begin{figure*}[t]
	\centering
	\subfigure[$\mu_i=0$]{\includegraphics[width=0.28 \linewidth]{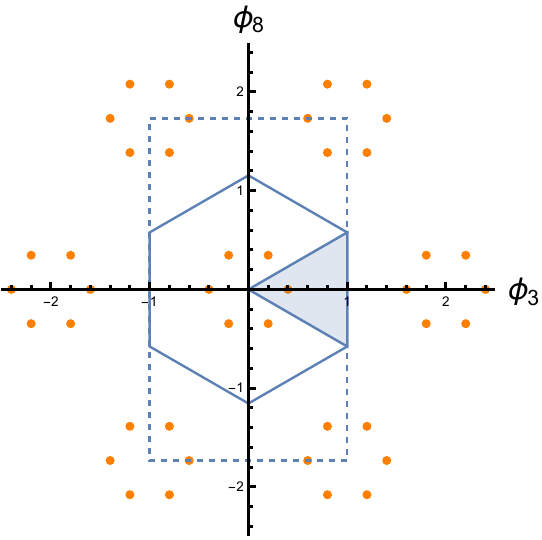}}
	\qquad
	\subfigure[$\mu_i=-2/3\pi T,\, \phi_8=\phi_8'+\frac{2\sqrt{3}}{3}$]{\includegraphics[width=0.28 \linewidth]{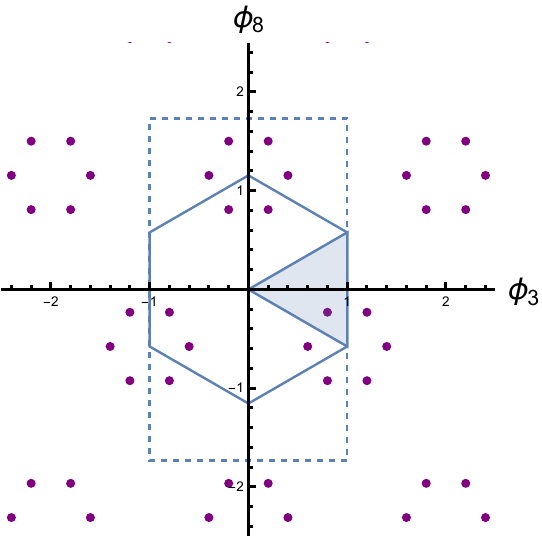}}
	\qquad
	\subfigure[$\mu_i=-4/3\pi T ,\, \phi_8=\phi_8'+\frac{4\sqrt{3}}{3}$]{\includegraphics[width=0.28 \linewidth]{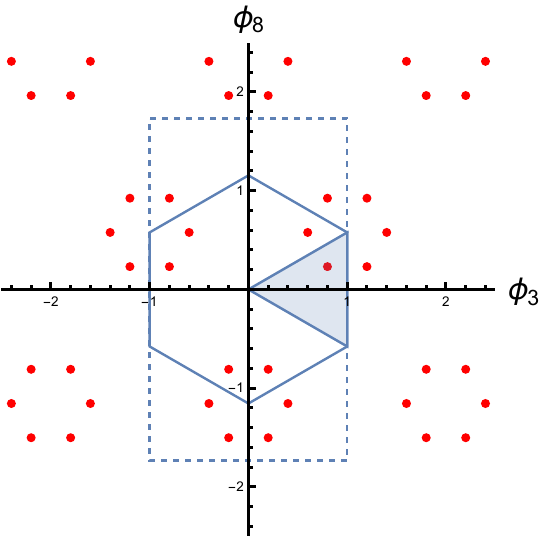}}
    \qquad
	\caption{The minima of Polyakov loop potential at specific imaginary chemical potential $\mu= -\frac{2n\pi T}{3} i, \ n \in \mathbb{Z}$. With a shift of $\mu_i$ by $-\frac{2}{3}\pi T$, the minima switch to the nearby center sector, and the Polyakov loop transforms as $L \to zL$, where $z = e^{2\pi i/3}$.}
	\label{fig:phi sym mui}
\end{figure*}

\begin{figure}[t]
	\centering
	\subfigure[Full Polyakov loop potential of QCD with finite quark mass at $T=0.15\gev$]{
		\label{fig:phi sym a}
        \includegraphics[width=0.46\linewidth]{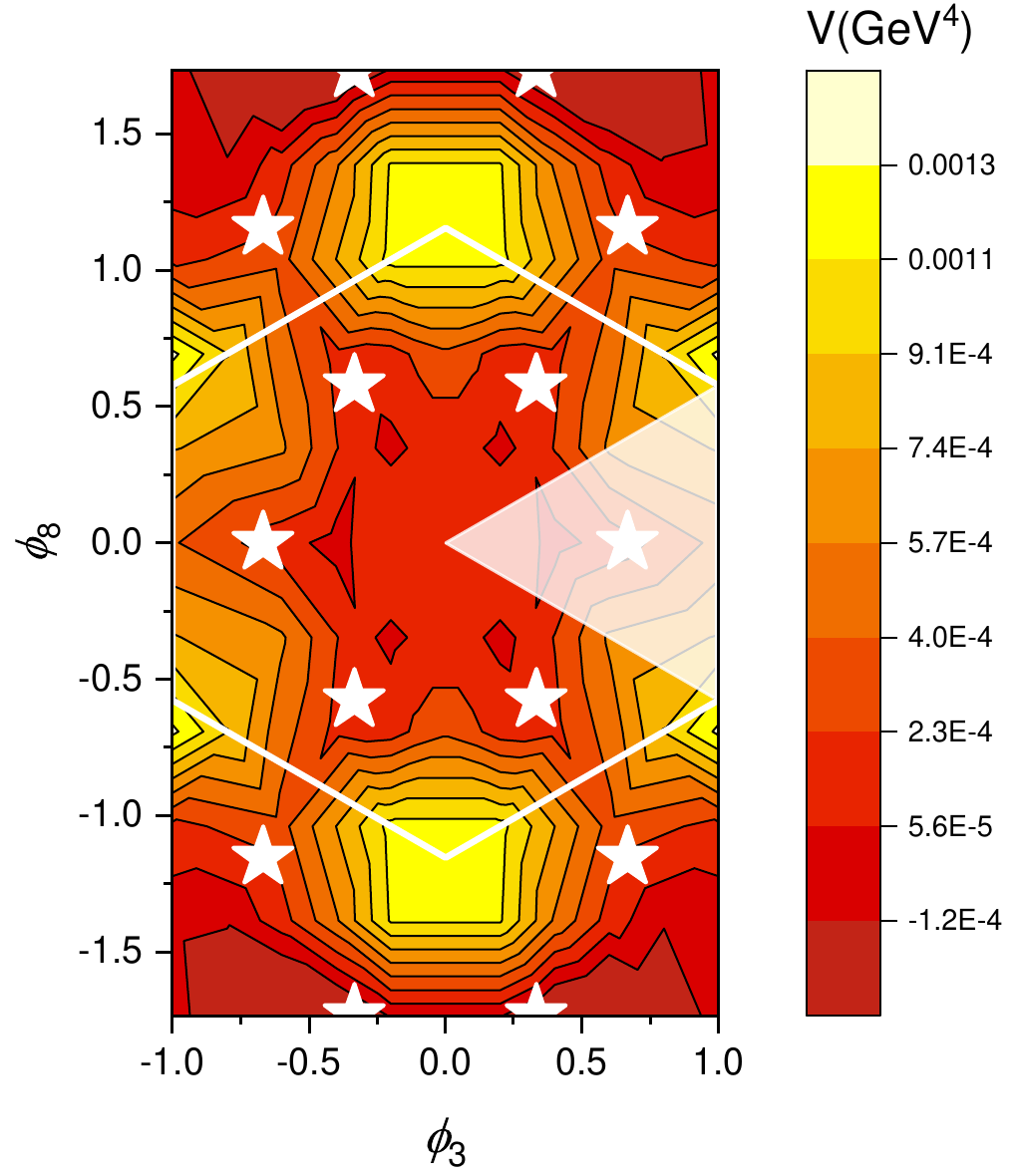}}
	\subfigure[Polyakov loop potential of only the gauge field part as an approximate "Yang-Mills" theory at $T=0.25\gev$]{
		\label{fig:phi sym b}
		\includegraphics[width=0.46 \linewidth]{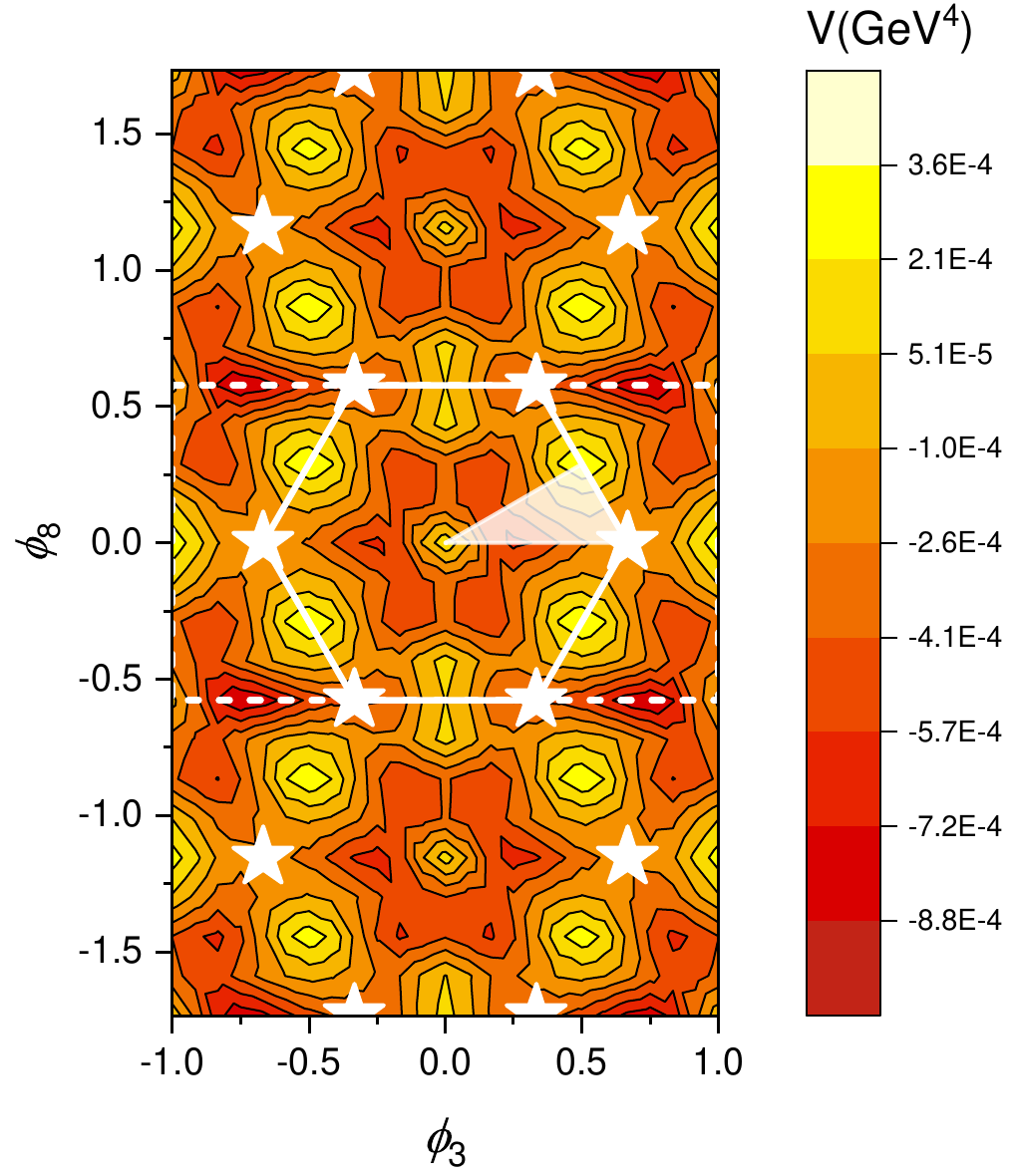}}
	
	\caption{The two panels display contour plots of the Polyakov loop potential $V$ for full QCD (left) and gauge field part in \eq{eq:Polya} (right), respectively. Dark red regions correspond to the potential minima, while stars denote the degenerate points where Polyakov loop $L=0$. 
    The hexagonal boundary and the shaded triangular region represent the elementary cell and fundamental domain of the symmetries, under a generic chemical potential.}
	\label{fig:phi sym}
\end{figure}

\begin{equation}\label{eq:symt2}
	(\phi_{3,r},\phi_{8,r})\rightarrow(\phi_{3,r}\pm\pi,\phi_{8,r}\pm\sqrt{3}\pi)
\end{equation}
\indent (c) The pair $(\phi_{3}, \phi_{8})$ possess rotation symmetry with rotation angle $\theta = \frac{\pi}{3}n$ for any integer $n$:
\begin{equation}\label{eq:symr}
(\phi_{3},\phi_{8})\rightarrow(\phi_{3}\cos \theta+\phi_8 \sin \theta,-\phi_{3}\sin \theta+\phi_8 \cos \theta)
\end{equation}
\indent (d) \label{sym:d} $(\phi_{3}, \phi_{8})$ represent charge conjugation invariance if $\mu=0$.
\begin{equation}\label{eq:symc}
	(\phi_{3},\phi_{8})\rightarrow(\pm\phi_3,\pm\phi_8).
\end{equation}
The first two symmetries originate from the periodicity of the real parts of $\phi_3$ and $\phi_8$, whereas the third symmetry arises from the global color symmetry. The Polyakov loop $L$ remains invariant under these transformations as well. These symmetry properties are illustrated in \fig{fig:phi sym a}, which shows the contour plot of QCD potential $V$ at temperature $T = 0.15~\text{GeV}$ and vanishing chemical potential ($\mu = 0$). The elementary cell and fundamental domain of these three symmetries are represented by a hexagonal boundary and a shaded triangular region, respectively. As shown in the figure, the degenerate minima of $V$ for the QCD system appear as dark red regions, distinctly separated from the white star markers indicating points where $L = 0$ and center symmetry is preserved. This separation between the potential minima and center symmetric points clearly demonstrates the explicit breaking of center symmetry in QCD systems with finite quark masses.

The same theoretical analysis applies equally to pure Yang-Mills theory, which shares all the aforementioned symmetry properties. Nevertheless, several crucial differences emerge. 
Notably, pure Yang-Mills theory consistently exhibits charge conjugation symmetry, which leads to the fundamental domain being reduced to half that of QCD theory. In addition, the translation period of $\phi_8$ is shortened to $2\pi\sqrt{3}/3$. This additional symmetry is referred to as the center symmetry. At low temperature, the minima of the Polyakov loop potential for pure Yang-Mills theory are located at the $(\phi_3,\phi_8)$ plane where the Polyakov loop $L=0$. As the temperature increases till the first-order phase transition region, The minima become the points with nonzero $L$, and consequently, there exists three degenerate minima, corresponding to $L \in \{l, zl, z^2l\}$ where $l \in \mathbb{R}$ and $z = e^{2\pi i/3}$. In another word, such a  triplet of minima signals the spontaneous breaking of the center symmetry in pure Yang-Mills theory. The characteristic contour plot at $T = 0.25~\gev$ for the gauge field part of QCD, where this phase transition occurs, is presented in \fig{fig:phi sym b}. The gauge field part has the same symmetry as the pure Yang-Mills theory, but the location of the phase transition may differ.  
%

\subsection{Roberge-Weiss symmetry}
The imaginary chemical potential is more than just an analytic extension of the real one; it carries abundant symmetry. The center transformation $z$ acting on the Polyakov loop $L$ is equivalent to some specific values of imaginary chemical potential given by:
\begin{equation}\label{eq:center2}
	\mu_i = -\pi T \theta = \frac{2\pi T}{3} n, \quad n \in \mathbb{Z}
\end{equation}
In this context, the minima of $(\phi_3, \phi_8)$ shift to another center sector, like $\phi_8 = \phi'_8 + \sqrt{3}\theta$. Consequently, \eq{eq:wn} transforms into the following forms:
\begin{equation}
	\begin{aligned}
		& \omega_n + i \mu_r + \phi_3 \pi T - \phi'_8 \frac{1}{\sqrt{3}} \pi T, \\
		& \omega_n + i \mu_r - \phi_3 \pi T - \phi'_8 \frac{1}{\sqrt{3}} \pi T, \\
		&\omega_n + i \mu_r + 3\theta \pi T + \phi'_8 \frac{2}{\sqrt{3}} \pi T.
	\end{aligned}
\end{equation}
Obviously, the quark propagator remains invariant if $\theta$ takes a value in \eq{eq:center2}, leading to the invariance symmetry in thermodynamic quantities, known as the Roberge-Weiss symmetry \cite{Roberge:1986mm}. However, the minima of Polyakov loop potential shift as illustrated in \fig{fig:phi sym mui}.

 Additionally, as $\mu_i'=\mu_i-2/3\pi T$ the Polyakov loop acquires a phase factor, $L \rightarrow zL,\ z=e^{2/3\pi i}$. According to \cite{Roberge:1986mm}, $L$ does not always undergo a crossover at the midpoint of one Roberge-Weiss period, given by $\mu_i = (1 + 2n) / 3\pi T$ for $n \in \mathbb{Z}$. In fact, at high temperatures, when $L$ is sufficiently large, it undergoes a first-order phase transition. The endpoint of the RW line, which has been extensively discussed in various works \cite{Kouno:2009bm, Scheffler:2011te, Cuteri:2022vwk} remains a topic of debate concerning whether it represents a second-order transition or a tricritical point.  These transitions are referred to as Roberge-Weiss transitions, and the midpoint line is known as the RW transition line.
 %
 
%
\section{results and discussions}\label{results}
Below, we present our findings concerning the QCD phase transition, the RW transition, and LYES under varying chemical potential conditions. Our computations reveal that, when $\mu = 0$, the minimum of Polyakov loop potential in the fundamental domain resides on the line spanned by $\phi_{3,r}$. Conversely, for $\mu \neq 0$, \tab{tab:muphi} summarizes the variables that require scanning. Further details are provided in the subsequent subsections. 
\begin{table}[htbp]
\caption{For specific chemical potential $\mu$, we find the minima of the Polyakov loop potential in complex $(\phi_3,\phi_8)$ plane. Within the fundamental domain shown in \fig{fig:phi sym a}, only the $\phi$ variables marked with $\checkmark$ need to be determined. We show all other variables are zero.}
	\begin{tabular}{|l|c|c|c|c|}
		\hline
		&~~~ $\phi_{3,r}$~~~ &~~~ $\phi_{3,i}$~~~ & ~~~$\phi_{8,r}$~~~ &~~~ $\phi_{8,i}$~~~	\\
		\hline
		$\mu=0$ & $\checkmark$ & 0 & 0 & 0 \\
		\hline
		$\mu_r\neq 0,\,\mu_i=0$ & $\checkmark$  & 0 & 0 & $\checkmark$ \\
		\hline
		$\mu_i\neq 0,\,\mu_r=0$ & $\checkmark$  & 0 & $\checkmark$ & 0 \\
		\hline
		$\mu_{r,i}\neq 0$ & $\checkmark$  & $\checkmark$ & $\checkmark$ & $\checkmark$ \\
		\hline
	\end{tabular}
    \label{tab:muphi}
\end{table} 
\subsection{At finite temperature and real chemical potential}
At zero chemical potential, considering $\phi_{3,r}$ alone is sufficient to pinpoint the minima of the system. Several points are marked with green dots in \fig{fig:Mlist}. It is evident that, within the temperature range $T \in (0.1, 0.17) \, \text{GeV}$, $\phi_{3,r}$ significantly impacts the chiral symmetry, exacerbating the breaking of this symmetry.

Furthermore, we direct our attention to the QCD phase transition at finite temperatures. We proceed to compute the chiral condensate $\langle \bar{\psi} \psi \rangle$ from the quark propagator using the following equation:
\begin{equation}\label{con}
	\langle \bar{\psi} \psi \rangle = -Z_{2}Z_{m} N_{c} T \sum_{n} \int \frac{d^{3}p}{(2\pi)^{3}} \, \text{tr}[S(\tilde{\omega}_{n}, p)] \, ,
\end{equation}
With a finite quark mass, the chiral condensate needs to be regularized, we take use of the reduced condensate $\langle \bar{\psi} \psi \rangle_{\text{red}}$ in Ref.~\cite{Gao:2021wun}:
\begin{equation}\label{eq:cond}
	\langle \bar{\psi} \psi \rangle_{\mathrm{reg}} = \langle \bar{\psi} \psi \rangle_l - m\frac{\partial}{\partial m}\langle \bar{\psi} \psi \rangle_l .
\end{equation}
And the light-quark chiral susceptibility $\chi_m^l$ in Ref.~\cite{Fischer:2014ata} is  
\begin{equation}\label{eq:chim}
	\chi_{m}^{l} = \frac{\partial \langle \bar{\psi} \psi \rangle_{l}}{\partial m_{l}} \,.
\end{equation}

To determine the pseudo-critical temperatures associated with chiral symmetry restoration and the transition to deconfinement, we use the two order parameters: the maximum value of the light-quark chiral susceptibility $\chi_m^l$, and the inflection point of the derivative of the Polyakov loop $\frac{dL}{dT}$. We obtain the pseudo-critical temperature for chiral symmetry restoration is $T_\chi = 0.162 \,\gev$, and for confinement transition is $T_{\text{conf}} = 0.146 \,\gev$. It is noteworthy that the presence of a non-zero value of $\phi_{3,r}$ raises $T_\chi$ from the previously reported value of $0.155 \, \gev$ in Ref.~\cite{Wan:2024xeu}.
 \begin{figure}[htbp]\centering
  \includegraphics[width=0.9\linewidth]{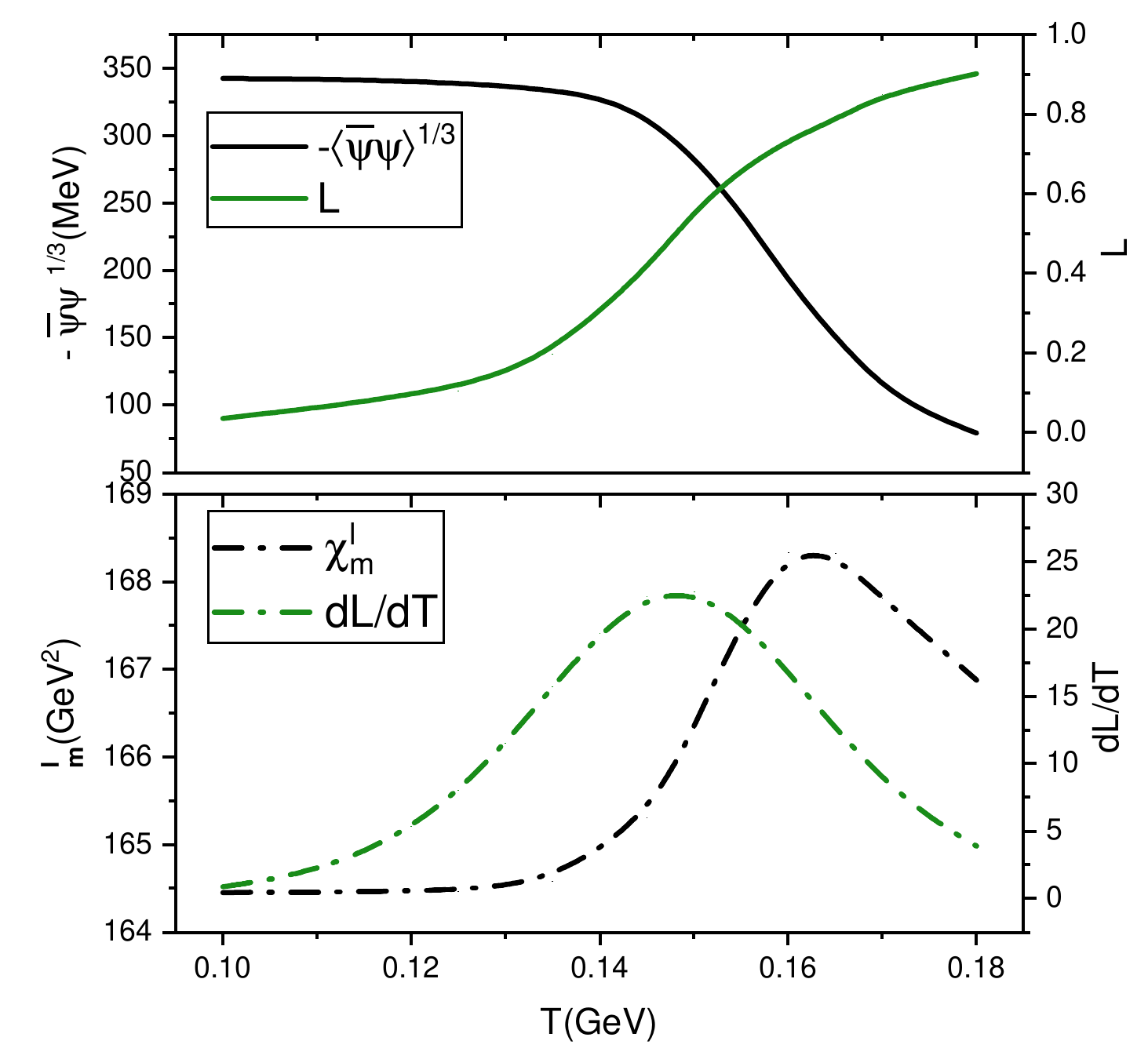}
  \caption{The behavior of the chiral condensate $\langle \bar{\psi} \psi \rangle$ and the Polyakov loop $L$ in the presence of a background gluon field.}\label{fig:LM}
	 \end{figure}

\begin{figure}[htbp]
\includegraphics[width=0.8 \linewidth]{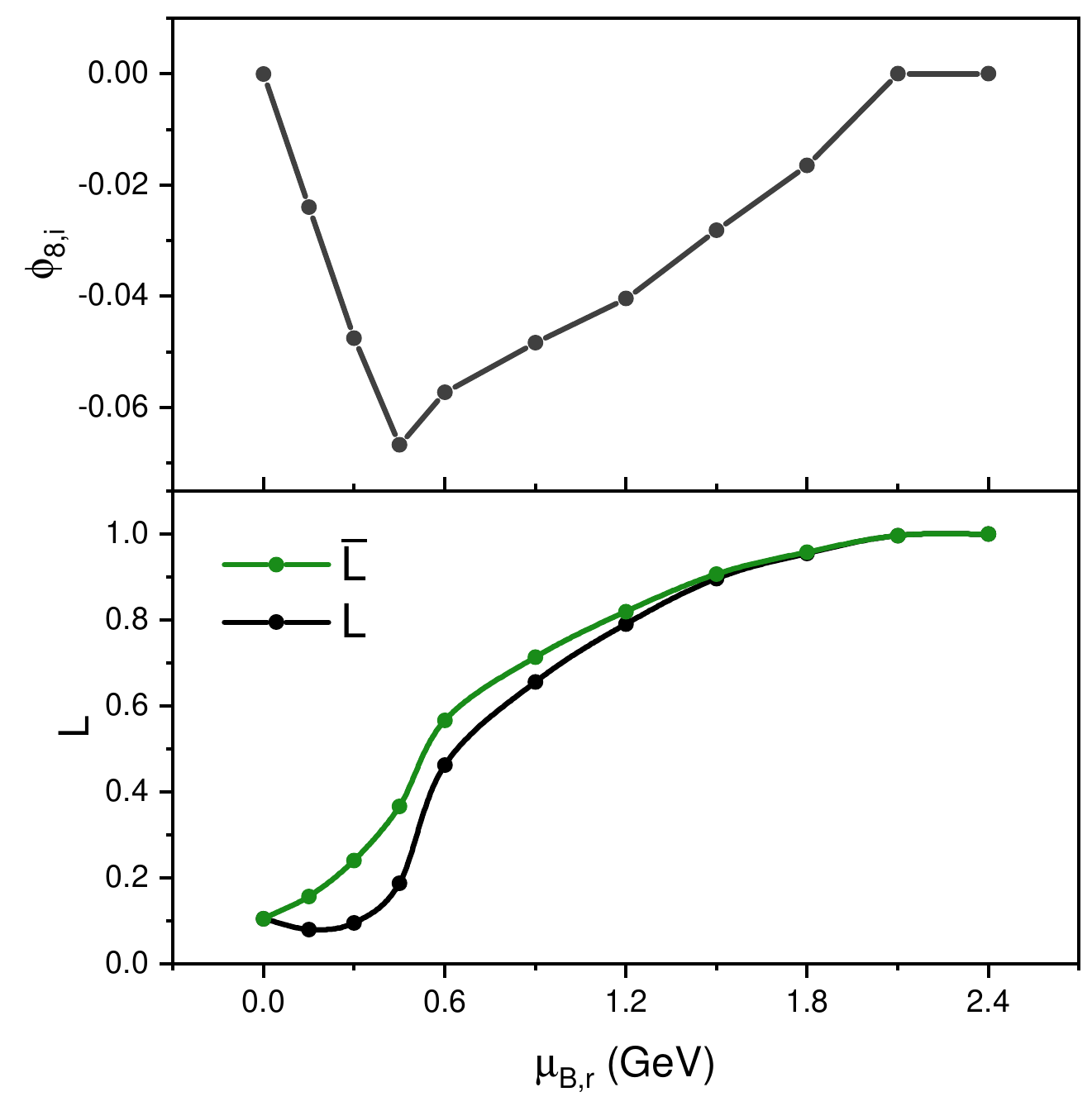}
\caption{$\phi_{8,i}$ and $L$,$\bar{L}$ with respect to real chemical potential at $T=0.125\gev$.}
\label{fig:mur}
\end{figure}

At non-zero real chemical potential,  the behavior of  $A_4$ condensate is similar, however, due to the analytic property and Cauchy-Riemann theorem, the imagninary of $\phi_8$ is required. The minima of the system are found to lie within the $(\phi_{3,r}, \phi_{8,i})$ plane. \fig{fig:mur} illustrates the behavior of $\phi_{8,i}$ at a temperature $T = 0.125$ GeV as a function of $\mu_r$. Initially, $\phi_{8,i}$ starts at zero and decreases to negative values as $\mu_r$ increases. Subsequently, as $\mu_r$ continues to rise, $\phi_{8,i}$ increases back to zero.
As already indicated in Table~\ref{tab:muphi}, non-zero values of $\phi_8$ are induced from non-zero $\mu$, which suggests the relationship between $\phi_8$ and charge conjugation. More specifically, this relationship can be inferred from the expressions involving the Polyakov loop $L$ and its conjugate $\bar{L}$ as follows:
\begin{equation}
	\begin{split}
		L(\bar{A}_4 )=
		& e^{i\int \bar{A}_4 \text{d}x^4}=e^{-F_q/T}\\
		\bar{L}( \bar{A}_4  ) =&e^{-i\int \bar{A}_4 \text{d}x^4}=e^{-F_{\bar{q}}/T}
	\end{split}
\end{equation}
As $L$ is an even function of $\phi_3$ shown in \eq{eq:L}, the difference between $L$ and $\bar{L}$ is manifested through non-zero values of $\phi_8$. This $\phi_8$ represents the disparity in free energy between static quarks and anti-quarks, which manifests the degree of charge conjugation asymmetry as illustrated in Sec.~\ref{sym:d}.d. Initially, the absolute value of $\phi_8$ increases as $\mu_r$ rises, introducing a distinction between quarks and anti-quarks. However, as $\mu_r$ continues to climb, the chiral symmetry starts to restore, leading both quarks and anti-quarks to exhibit \textit{Asymptotic Freedom}. Consequently, the discrepancy in their free energies gradually diminishes.
%
%
According to \eq{eq:L}, a negative value of $\phi_{8,i}$ implies that $\bar{L} > L$, as illustrated in \fig{fig:mur}. This indicates that the free energy of anti-quarks ($F_{\bar{q}}$) is less than that of quarks ($F_q$). This finding is in agreement with the results presented in \cite{Fukushima:2009dx}, which state that in a quark medium, the color screening effect on test anti-quarks is more pronounced, yielding a larger value of $\bar{L}$.

%
%
\subsection{The Lee Yang edge singularities of both RW and Chiral phase transitions}\label{LYE}
\begin{figure}
     \centering
     \includegraphics[width=0.8\linewidth]{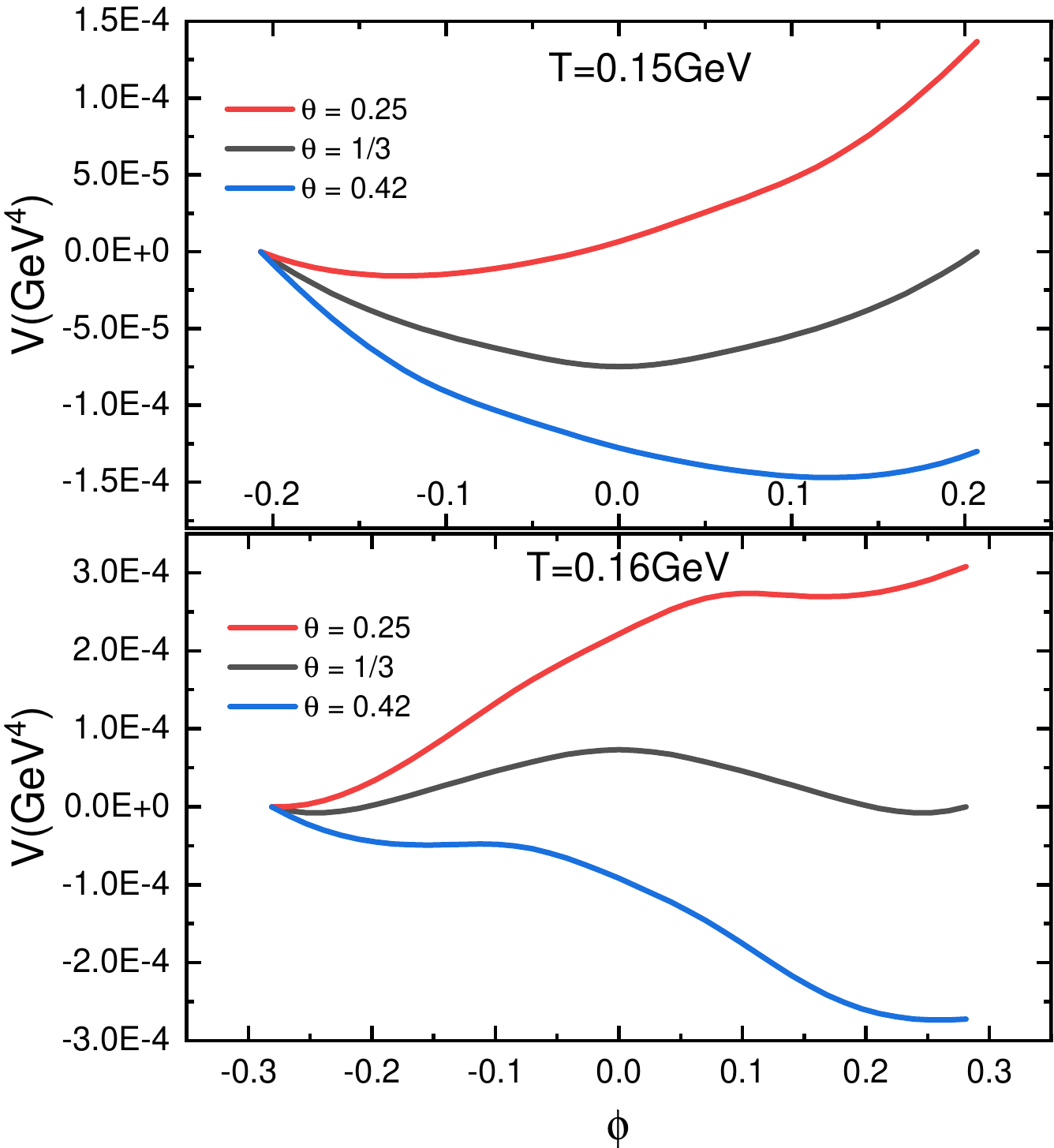}
     \caption{The Polyakov loop potential at temperatures $T=0.15, 0.16~\gev$ and different imaginary chemical potentials $\mu_i=-\pi T\theta$, along the respective direction defined in \eq{eq:defofphi} with the control parameter $\phi$. In particular at $\theta=1/3$, $\phi=0$ corresponds to the background field value $(\phi_{3,0},\phi_{8,0})$ for the extreme point of the potential which satisfies \eq{eq:PolyakovSym} and \eq{eq:minimalofphi}.
     }
     \label{fig:muipl}
 \end{figure}
To study the Lee Yang edge singularities, one needs to generalize the system into imaginary chemical potential. Straightforwardly, a new RW symmetry appears in the $(\phi_3,\phi_8)$ plane, and induces a new phase transition. The relation between this new phase transition and chiral phase transition is thus worthy of investigation. 

First of all, the Polyakov loop become complex for complex chemical potential, and hence, it is not appropriate to regard it as the usual deconfinement order parameter. The behavior of Polyakov loop at complex chemical potential together with the relation between deconfinement and the RW phase transition is beyond the scope of this paper, and here we directly consider the minimum of Polaykov loop potential in $(\phi_3, \phi_8)$ plane as the order parameter of RW phase transition.  

Generally speaking, we find that the RW phase transition occurs above a critical temperature. Above the critical temperature, it is a first order phase transition, while below the critical temperature, the RW phase transition become a crossover.  In particular, at zero chemical potential, the RW phase transition  takes place above the temperature $T=155$ MeV. As depicted in \fig{fig:muipl}, for the temperature above, a first order phase transition occurs at the imaginary chemical potential $\mu_i=-\frac{1}{3}\pi T$, while for the temperature below, the minimum in $(\phi_3, \phi_8)$ plane changes smoothly with respect to the imaginary chemical potential. 

\begin{figure}
    \centering
    \includegraphics[width=0.9\linewidth]{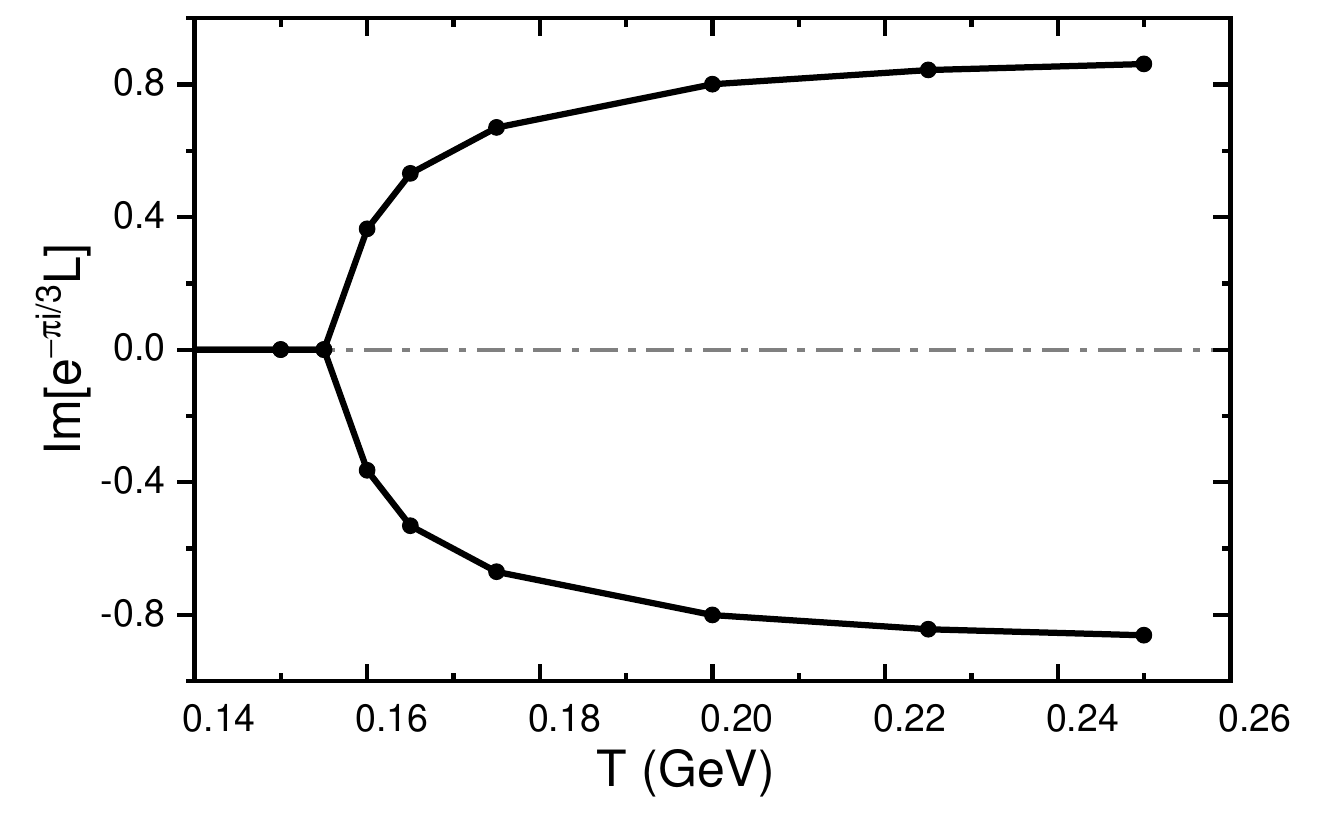}
    \caption{Temperature dependence of the imaginary part of Polyakov loop $L$ under  $e^{-\pi i/3}$ phase rotation.}
    \label{fig:RW_pt}
\end{figure}

To elucidate the connection between the Roberge-Weiss (RW) transition and center symmetry, let us analyze the QCD Lagrangian under an imaginary chemical potential $\mu_i = -\frac{1}{3}\pi T$:
\begin{equation}\label{eq:LagRW1}
    \mathcal{L} = \bar{\psi}\left(i\partial\llap{/} + g A_\mu\gamma^\mu + \mu_r\gamma^0 - i\frac{\pi T}{3}\gamma^0 - m\right)\psi - \frac{1}{4}F^a_{\mu\nu}F_a^{\mu\nu}.
\end{equation}
Applying a center transformation $z \in \mathbb{Z}_3$ to the Lagrangian, it transforms into its conjugate form:
\begin{equation}\label{eq:LagRW2}
    \mathcal{L}' = \bar{\psi}\left(i\partial\llap{/} + g A_\mu\gamma^\mu + \mu_r\gamma^0 + i\frac{\pi T}{3}\gamma^0 - m\right)\psi - \frac{1}{4}F^a_{\mu\nu}F_a^{\mu\nu}.
\end{equation}
If this symmetry is preserved in the physical states, the Polyakov loop $L$ must satisfy:
\begin{equation}\label{eq:PolyakovSym}
    L^* = z^{-1} L = e^{-2\pi i/3}L,
\end{equation}
implying $L = e^{\pi i/3}|L|$. This corresponds to: 
\begin{align}\label{eq:minimalofphi}
\phi_8 = \tan\left(-\frac{2\pi}{3}\right)\left(\phi_3 - \frac{2}{3}\right), \quad \phi_3\in[1/2,2/3]\,,   
\end{align}
in the fundamental domain of \fig{fig:phi sym a}. 
Following this idea, we further check directly from the physical point position of the Polyakov potential, whether the symmetry presented by \eq{eq:PolyakovSym} is broken or not. To do this, we first determine the extreme point of the potential along \eq{eq:minimalofphi} at $\theta=1/3$, which we denote it as $(\phi_{3,0}, \phi_{8,0})$.
Then, we extract the shape of potential at different $T$ and $\theta$, along a direction perpendicular to \eq{eq:minimalofphi}:
\begin{align}\label{eq:defofphi}
 (\phi_3,\phi_8) = ( \phi_{3,0} + \frac{\sqrt{3}}{2} \phi, \, \phi_{8,0} - \frac{1}{2} \phi ),
\end{align}
with $\phi$ as a control parameter.
Finally, we check whether the physical point, i.e. minimum of the potential indeed lies at $\phi=0$.
This is illustrated in Fig.~\ref{fig:muipl}.
We found that below the critical temperature $T_c = 0.155\gev$ at $\theta = \frac{1}{3}$, the minimum of the potential resides at $\arg(L) = \pi/3$, signaling unbroken symmetry behavior. However, when $T > T_c$, the minimum shifts to $\arg(L) \neq \pi/3$, indicating spontaneous breaking behavior. The complete temperature dependence of the imaginary part of Polyakov loop with a $e^{-\pi i/3}$ phase rotation: $\text{Im}(e^{-\pi i/3}L)$ is shown in \fig{fig:RW_pt}.

Therefore,  for every temperature and chemical potential where the RW phase transition can happen, it happens at a certain imaginary chemical potential as:
\begin{equation}
\mu_i=(1 + 2n) / 3\pi T,
\ n\in \mathbb{Z}. 
\end{equation}

\begin{figure}[t]
	\centering
        \subfigure[The LYES trajectories of QCD phase transitions in the $\re \, \mu_B/T$ - $\im \, \mu_B/T$ plane.]{
		\label{fig:lye traj}
		\includegraphics[width=0.8 \linewidth]{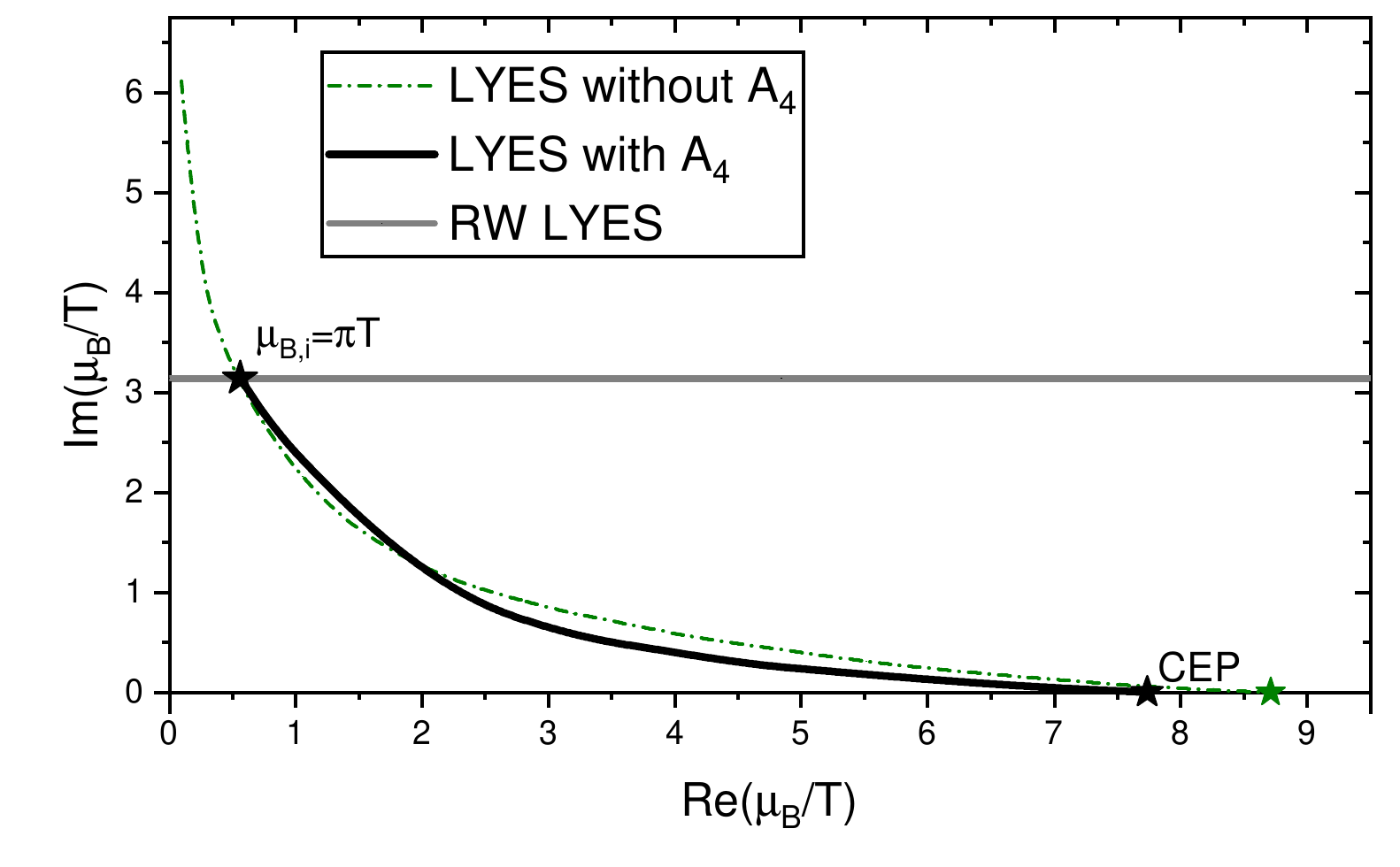}} \\
	\subfigure[The LYES trajectories in the vicinity of CEP location, and the respective scaling fits according to \eq{eq:cep_scal} with $\beta\delta$ and the slope parameter $c_2$ in GeV unit.]{
		\label{fig:cep scal}
        \includegraphics[width=0.8\linewidth]{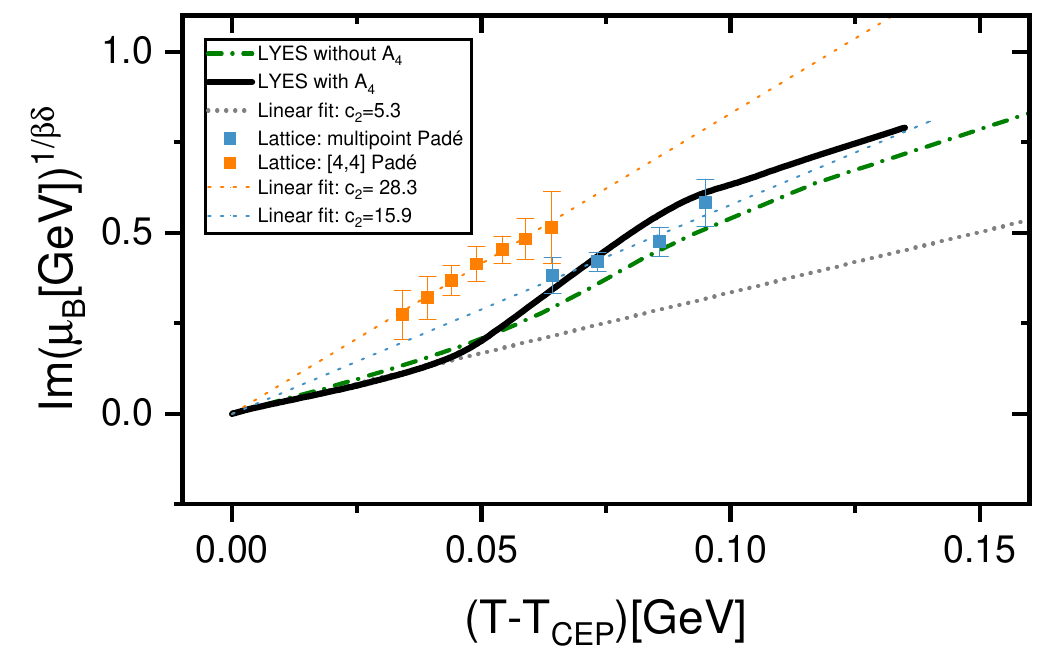}}\\
	\caption{Upper panel: the LYES trajectory with (black-solid) and without $A_4$ (green-dotted). The background field does not affect the critical exponent beyond the error margin, however, it generates a cut-off temperature at $T_e=0.235$ GeV that above this temperature there exists no LYES due to the RW phase transition. Lower panel: scaling analysis on the trajectories in the vinicity of the CEP location: a linear fit (grey-dotted line) on the trajectory with $A_4$ is performed according to~\eq{eq:cep_scal}, which yields $\beta\delta = 1.38$ and $c_2=5.3$ (GeV unit). We also show the LYES extracted from lattice QCD simulations using the multi-point Pad\'{e}~\cite{Dimopoulos:2021vrk} and the [4,4] Pad\'{e}~\cite{Bollweg:2022rps} approaches, together with their scaling fits~(blue-dashed and orange dashed lines) to the extrapolated CEP locations.}
	\label{fig:LYESs} 
\end{figure}

Such a transition brings in additional singularities in imaginary chemical potential region, which affects the behavior of  the LYES in QCD.
We then replicate the methodology outlined in \cite{Wan:2024xeu} within the current framework to investigate the impact of $\bar{A}_4$ on  Lee-Yang edge singularities (LYES). The trajectories of the LYE singularities for both schemes are depicted in \fig{fig:LYESs} for comparison.


One significant outcome of incorporating $\bar{A}_4$ is that the LYE singularities  has represented a cut-off temperature, at $T_e = 0.235$ GeV. This is because at this temperature, the imaginary part of the LYE singularity  approaches 
the  Roberge-Weiss transition line, given by $\mu_{\lye,i} = \frac{1}{3}\pi T$. As the temperature increases, in the original scheme, $\mu_{\lye,i}$ would grow larger. However, in the current  scheme,  the RW symmetry entails that $\bar{A}_4^a$ is symmetric to  the axis $\mu_{i}=\frac{1}{3}\pi T$. Therefore, the values of $\bar{A}_4$ and $\mu_i$ at $\mu_{i} > \frac{1}{3}\pi T$ leads to a singularity that is located  at $\mu_{i} < \frac{1}{3}\pi T$, and hence, it will never reach the singularity in the correlation function at $\mu_{i} > \frac{1}{3}\pi T$. Consequently, the LYE singularities at higher temperatures terminate within the branch cut of the Roberge-Weiss first-order phase transition. 

As depicted in \fig{fig:LYESs},  the consideration of the gluonic background $A_4$ condensate  does not change greatly of the behaivor of LYES below $\mu_i=\frac{1}{3}\pi T$ and $T_e=0.235$ GeV. For increasing temperature  above critical end point~(CEP): $(T_{\textrm{CEP}},\mu_{B,\textrm{CEP}})=(0.095,0.735)~\gev$, the location of LYES is monotonously changing to smaller $\mu_r$ and larger $\mu_i$. However, the inclusion of the $A_4$ condensate prevents the LYES for larger temperature, and makes the further LYES fixed at the RW phase transition line.  {The termination of LYES at $T_e$ also makes the extrapolation of CEP location  invalid  with the LYES near the temperature $T\approx0.235$ GeV.

Furthermore, we verify the CEP scaling for our current results, finding that the critical exponents $\beta\delta$,  representing the relationship between the imaginary part of the LYES and the temperature 
\begin{equation}
	\text{Im} \mu_{LYE}=i c_2 (T-T_{\text{CEP}})^{\beta\delta}\,,
    \label{eq:cep_scal}
\end{equation}
 is unchanged with the existence of  $\bar{A}_4$. The points of $(\text{Im} \mu_{LYE})^{1/\beta\delta}$ with respect to $T-T_{\text{CEP}}$ of our previous study~\cite{Wan:2024xeu} and current one are shown in \fig{fig:cep scal}. Within the scaling region $T - T_{\text{CEP}} \lesssim 0.04 \, \text{GeV}$, taking $\beta\delta = 1.38$, both  datasets exhibit linear behavior, demonstrating that 
 the critical behaviour of the LYES is not sensitive to the presence of a confining gluon background field. It is then intuitive to think that the chiral dynamics dominates in the respective region.
 We obtain $c_2=5.3$ GeV$^{1-\beta\delta}$ and the dimensionless slope parameter $\bar{c}_2=c_2\frac{(T_{\text{CEP}})^{\beta\delta}}{\mu_{B,\text{CEP}}}=0.28$, which is consistent with our previous result for the case without $\bar{A}_4$. Following the same analysis of CEP scaling in~\cite{Wan:2024xeu}, this yields an extracted CEP location from the LYES trajectory at $(T,\mu_B)=(118,608)$ MeV.
Moreover, based on our results, one may expect that if using the LYES with the temperature for $T\leq$ 160 MeV,  the extrapolation with the CEP scaling is  plausible for determining the CEP location of QCD.


\section{Summary}\label{sum}
In this paper, we have employed the Dyson-Schwinger Equation (DSE) approach for Quantum Chromodynamics (QCD) to investigate the quark propagator at complex chemical potential in the presence of a background gluon field condensate. Our primary focus has been on studying the QCD phase transition. In the context of a background gluon field, the behavior of the chiral condensate, denoted as $\langle \bar{\psi} \psi \rangle$, and the Polyakov loop, $L$, exhibits intriguing characteristics. The chiral condensate, which serves as an order parameter for chiral symmetry breaking, undergoes significant modifications due to interactions with the gluon field. Concurrently, the Polyakov loop, which is indicative of deconfinement phases, also displays a notable response to the presence of the gluon field background.

We summarized the main results concerning the DSEs with $A_4$ condensate, illustrating how the inclusion of such a field affects the momentum translation of Matsubara frequencies for different color components. This analysis laid the groundwork for deriving the quark propagator and bare gluon propagator in the presence of the background field. Such  scheme allowed for a comparative assessment of the background field's impact on both the chiral phase transition and LYES. We presented the results for the Roberge-Weiss critical point for light quarks and its associated LYES in the low-temperature region. Furthermore, we explored the chiral phase transition LYES while accounting for a constant background gluon field, specifically the $A_4$ component, which is crucial for understanding confinement and center symmetry.
We examined the trajectories of LYES, finding that the inclusion of the background field affects the location of these singularities but does not significantly alter the critical exponent.

We also delved into the symmetries of the $(\phi_3, \phi_8)$ plane, which provided insights into the periodic and rotational symmetries of the Polyakov loop potential. These symmetries were crucial in simplifying the search for minimum points of the potential and understanding the behavior of the Polyakov loop under various conditions. Our results indicated that the presence of a finite quark mass explicitly breaks center symmetry, leading to a deviation of the minimum points from the center of the symmetry plane as temperature increases.

Overall, this study contributes to a deeper understanding of the QCD phase transition at complex chemical potential, highlighting the importance of considering background gluon fields and their impact on chiral symmetry breaking and confinement.


\medskip

\section{Acknowledgements}
The authors give special thanks to Friederike Ihssen and Jan M. Pawlowski for discussions. 
YL and FG also thank the other members of the  fQCD collaboration~\cite{fQCD} for discussions.
This work  is supported by the National Natural Science Foundation of China under Grants  No. 12247107, No. 12175007. FG is   supported by the National  Science Foundation of China under Grants  No. 12305134.

\bibliography{reference}

\end{document}